# RESEARCH

# iGPSe: A Visual Analytic System for Integrative Genomic Based Cancer Patient Stratification

Hao Ding[1], Chao Wang[3], Kun Huang[2] and Raghu Machiraju[1*]

## Abstract

**Background:** Cancers are highly heterogeneous with different subtypes. These subtypes often possess different genetic variants, present different pathological phenotypes, and most importantly, show various clinical outcomes such as varied prognosis and response to treatment and likelihood for recurrence and metastasis. Recently, integrative genomics (or panomics) approaches are often adopted with the goal of combining multiple types of omics data to identify integrative biomarkers for stratification of patients into groups with different clinical outcomes.

**Results:** In this paper we present a visual analytic system called Interactive Genomics Patient Stratification explorer (iGPSe) which significantly reduces the computing burden for biomedical researchers in the process of exploring complicated integrative genomics data. Our system integrates unsupervised clustering with graph and parallel sets visualization and allows direct comparison of clinical outcomes via survival analysis. Using a breast cancer dataset obtained from the The Cancer Genome Atlas (TCGA) project, we are able to quickly explore different combinations of gene expression (mRNA) and microRNA features and identify potential combined markers for survival prediction.

**Conclusions:** Visualization plays an important role in the process of stratifying given population patients. Visual tools allowed for the selection of possibly features across various datasets for the given patient population. We essentially made a case for visualization for a very important problem in translational informatics.

## Background

During the past fifteen years, high-throughput genomic experiments, which involve the usage of microarrays or next generation sequencing technologies, have significantly changed biomedical research and clinical practice. These technologies have expedited the process of discovering genes implicated in important biological phenomena and molecular markers for disease. However, for a better understanding of large high-dimensional data associated with high-throughput experiments, intuitive visualization tools are needed in order to effectively interpret the data and extract new biological knowledge and insights. Specifically, there is an important and unmet need to integrate collected data with previous biological and clinical knowledge. It is thus important to access large repositories of often structured biological knowledge and further to have access to interactive tools that facilitate the required integration and analysis of all data, especially against the backdrop of prior knowledge. The specific need of integrative visual analytics is of particular importance given the growing trend of using integrative genomics (or panomics) approaches for personalized treatment of diseases, including cancer.

Most types of cancers are highly heterogeneous with different identifiable subtypes. These subtypes often possess different genetic variants, present different pathological phenotypes, and most importantly, confer different clinical outcomes such as varied prognosis and response to treatment as well as different likelihood for recurrence and metastasis. Patient stratification is necessary for prescribing viable regimens of treatment and also towards the discovery of prognostic and/or predictive biomarkers.

For instance, there are four main subtypes of breast cancer: Luminal A (LumA), Luminal B (LumB) and Triple Negative/Basal-like, and HER2-type. The luminal subtypes are associated with the expression of estrogen receptor (ER) and progesterone receptor (PR), while HER2-type usually lacks hormone receptor expression (ie, ER- and PR-) but have amplification

[*]Correspondence: raghu@cse.ohio-state.edu
[1]Computer Science& Engineering Department, The Ohio State University, 43210 Columbus, OH, US
Full list of author information is available at the end of the article and/or over-expression of HER2. Basal-like tumors are commonly described as triple-negative breast cancers (TNBCs) lacking in expression of hormone receptors and the oncogene HER2 (HER2-). In order to robustly characterize patient subtype demographics to achieve precision medicine, panomics approaches are being increasingly used especially in tandem with the development of large collaborative projects such as TCGA (The Cancer Genome Atlas). In these projects, large cohorts of patients were recruited and many different types of "omics" data including genotypes (e.g., single nucleotide polymorphism), copy number variance, and somatic mutations), gene/microRNA expression, epigenomics (e.g., DNA methylation), proteomics, pathology images and clinical records as well as outcome information were collected from these patients. It is conceivable that by integrating the data ranging from genotype to multiple levels of phenotypes, more precise and robust stratification of the patients with clinical outcome difference can be achieved. Very least, conflicting stratifications arising from the consideration of each data collection separately are avoided, leading to identification of potentially more robust biomarkers.

However, integrative analysis is a challenging task. Most of such analysis requires extensive analysis and development of complicated algorithms such as network integration [1], statistical association and regression [2], and partial least square analysis [3]. Since many of the algorithms are still in the development and testing stage, users are often required to have extensive preparation in quantitative analysis, algorithm development and computational methods. These requirements severely hinder the wide utilization of such data by clinicians and biomedical researchers who are often only trained in clinical and biological sciences. Therefore, there is an urgent need for effective tools that allow biomedical researchers achieve tangible exploration of patient population using multiple types of "omics" and patient outcome data.

Here we present a visual analytic system called interactive Genomics Patient Stratification explorer (iGPSe), designed to help biomedical researchers to perform patient stratification on-the-fly and visually explore disease subtypes in heterogeneous genomic data. In iGPSe, we employ machine learning algorithms that identify sub-populations based on molecular feature sets chosen by users. To be effective, iGPSe relies on an interactive realization of parallel sets and patient survival plots between selected patient groups. Thus, investigators are able to quickly examine how a selected list of molecular features of interest can separate patients into different groups and whether these groups show a difference in clinical outcomes. In addition, iGPSe offers several visualization techniques to help users evaluate the quality of the stratification results and thus assess the effects of clustering algorithms. In this work, we demonstrate the use of iGPSe on the stratification of breast cancer patients using both mRNA and microRNA (miRNA) expression data. The most distinctive feature of iGPSe is a novel visualization scheme to explore combinations of different data types and identify combined markers for robust survival prediction of a given population. Additionally, iGPSe integrates standard patient stratification workflows into an intuitive, user friendly interactive platform.

Related work
Cancer tumors that seem very similar when examined through conventional diagnostic methods might look different at the molecular level leading to different and effective outcomes and/or treatment responses. Therefore, molecular features are being increasingly used to stratify patients to support more accurate and robust clinical and therapeutic decisions. Over the past decade, molecular stratification of tumors using gene expression microarrays has been an important area of cancer research [4, 5, 6, 7, 8, 9]. A typical stratification study often includes the application of statistical techniques to population groups including supervised learning and unsupervised cluster analysis. Heatmaps have been widely used in many tools [10] to visualize molecular signature patterns manifesting in various subgroups.

One of the popular genome tools is the UCSC Cancer Genomics Browser [11]. It allows researchers to identify and assess genomic signatures in cancer subtypes, to compare and contrast subtypes, and to assess their role in stratifying patients into different groups. However, this ability is restricted to one dataset at a time and offers no integrative capabilities. Many methods have been recently reported to discover characteristics from multiple classes of measurements [12, 13, 14, 15]. These computational methods either build statistical models [13, 14] or construct multiple networks or patient samples [15]. In bioinformatics, integrative analysis is becoming more prevalent with the increased adoption of the *integrative genomics* or the *panomics* [16, 17, 18, 19, 20]. A major challenge in any such integrative study is that the patient population is very heterogeneous for any given type of measurement data and thus the ensuing stratifications are often very different. Integrated visualization of heterogeneous data has not received much attention [21, 22].

A salient integrative visualization tool, StratomeX [23], was proposed towards the visualization and exploration of subtypes in a population afflicted with cancer using the TCGA data. StratomeX relies on on the visual acuity of heatmaps and the discriminative efficacy of clustering of collected genomics data. It is however difficult to discover the underlying structure that exists in a population and pose hypotheses that compares and contrasts characteristics and outcomes across sub-populations. Additionally, the high dimensionality of the data makes it difficult to display all of the relationships in a meaningful manner.

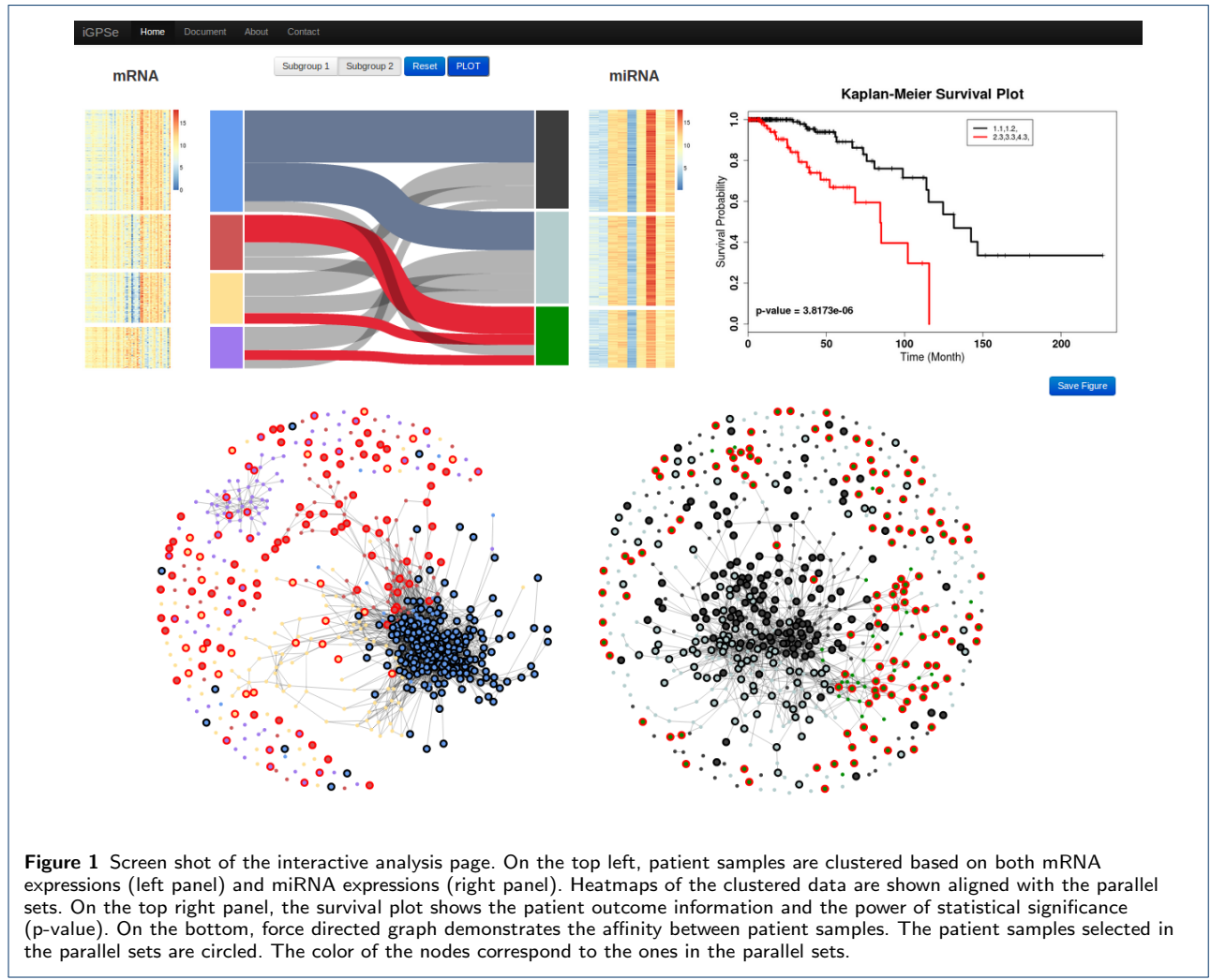

**Figure 1** Screen shot of the interactive analysis page. On the top left, patient samples are clustered based on both mRNA expressions (left panel) and miRNA expressions (right panel). Heatmaps of the clustered data are shown aligned with the parallel sets. On the top right panel, the survival plot shows the patient outcome information and the power of statistical significance (p-value). On the bottom, force directed graph demonstrates the affinity between patient samples. The patient samples selected in the parallel sets are circled. The color of the nodes correspond to the ones in the parallel sets.

## Methods

We developed the visual analytics system iGPSe, a web based tool facilitating patient stratification and exploration of disease subtypes in heterogeneous genomic data. We first elaborate the overall requirements of our system and the underlying design rationale. We continue by presenting visualization techniques adopted in iGPSe. We then describe the iGPSe workflow design and implementation.

Requirements Analysis and Design
In this subsection we outline various design decisions that ultimately guided our choices for computational methods and platforms. Our methods share commonalities with genome wide association studies (GWAS). To begin with, our studies will require a large population of patients. In addition, our collections of datasets are of multiple modalities (imaging, molecular expression, etc.), high-dimensional, large, and highly heterogeneous. In this section, we first describe input data under consideration and then enumerate various functionalities that will be required in a typical patient subtyping study. Eventually, we will make a strong case for the use of iGPSe to perform and evaluate patient stratification.

*Data Characteristics*
The population data used in iGPSe is extracted from the TCGA. For our current prototype, we limit our collection to include gene expression (mRNA) and micro-RNAs (miRNA) expression. The expression data were subject to log transformation and normalization following standard bioinformatics practice. Clinical information includes the age, tumor grade, survival time and survival status for each patient.

*User Wish List for iGPSe*
In order to identify the system requirements, we worked closely with domain experts. The suggestions and advise collected from our collaborators helped us to design, implement and improve our system. Accordingly, we identified the following functions, which are directly based on observations of the domain experts' standard data analysis workflows.

a) **Interactive feature selection:** Molecular features such as genetic variance and expression levels for selected genomic regions and genes are often used as potential biomarkers for identifying cancer subtypes, which helps in early diagnosis and effective treatment of cancer patients. Given the nature of cancer microarray data, which usually consists of a few hundred samples with thousands of genes as features, the selection of molecular features is important for effective gene expression data analysis. The feature selection step is a commonly addressed problem in machine learning especially in the context of supervised learning where different subtypes are labelled with prior domain knowledge. Thus, one can effectively eliminate the potentially irrelevant or obvious features. Users, therefore, should be able to add/drop/modify features interactively. Then by examining effects of eventual stratification, researchers can verify the quality of the selected feature set and refine it accordingly.

b) **Clustering:** It is now widely acknowledged that cancer is biologically heterogeneous. This complexity accounts partly for the variation in clinical outcome [16]. Given the large variations in genetic and environmental factors, there is a need to detect subpopulations and examine them under different annotations such as clinical outcome or histological types. One of the common methods is to cluster the patient population into subgroups of patients who share similar expression patterns. However, there is no one-size-fits-all solution to clustering. Each algorithm and similarity measurement impose certain assumptions on the data set. Different clustering algorithms can give widely differing stratifications, especially when deployed on gene expression data. Thus, the application should offer multiple choices of clustering algorithms for the user to choose. Moreover, the application should allow users to run clustering on-the-fly, rather than use precomputed results. In this way, users can choose the algorithm that works the best on their input data set and have more control over the stratification process. The details of our implementation is described in Section .

c) **Cluster refinement:** In many cases, the notion of a patient subgroups is not well defined in the selected feature space. The stratification of the population is, in most cases, not unique. Moreover, the number of the viable subgroups is also mostly unknown. It is important to do an evaluation of how well an algorithm performs on expression data sets. The application should provide some clues for the user to evaluate the quality of the clustering results, e.g, whether the chosen number of clusters is appropriate, the high-similarity within a cluster and the low-similarity between clusters. Visualization techniques can provide the evaluations. We integrated three visualization techniques into our system to help users analyze and improve the clustering results.

d) **Integrated analysis for patient stratification:** Our main goal is to provide various ways to form combinations of data types interactively. The visualization of these combinations should be intuitive and easy to navigate. The user must be able to detect patterns in the data through rearrangement and grouping. Further, he/she must be able to drill down to interesting details and obtain biological insight. It is also important to include all collected types of expression data [23]. Clustering results heavily depend on chosen parameters and different types of expression data will give rise to different clusters. Consequently the application should be able to compare stratification results from different data types, which could help researchers answer more complicated questions including whether there exist dependencies between stratification results from different data types. It should also offer interactive functions which help the user explore various combinations of subtypes. In order to perform the integrative analysis we adopt an interactive visualization method, parallel sets, which give an intuitive evaluation of the cluster coherence across different types of expression data and enable the user to interactively explore the various combinations of clustering. Details of our integrated analysis are described in Section .

e) **Interactive user interface:** Interactivity is an essential requirement given the complex nature of variation inherent in the datasets and the multitude of measurements and outcomes. The user needs to be able interactively to select sub-populations and clusters and compare outcomes and intrinsic labels between them. The user interface should be clear and intuitive, so that users with no computer science background are still able to operate the application. Moreover, it should be feasible to customize certain aspects of the visualization.

Visualization Components

In this section we describe the visualization components that are designed to address the aforementioned requirements. We adopted four visualizations techniques in our system. The heatmap helps users to examine the expression patterns in clusters. The silhouette plot provides an assessment of the relative quality of patient clustering. Graph visualization facilitates the exploration of the overall population in the selected feature space. The parallel sets display enables users to evaluate cluster coherence between two data modalities. Moreover, representing sub-populations of patients in parallel sets allows the user to interactively visualize the data, as well as to perform statistical tests to ascertain differences in clinical outcomes.

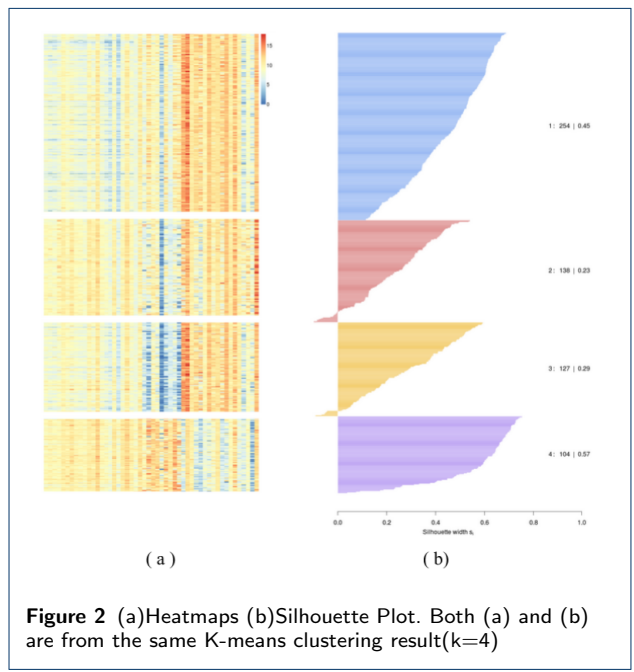

**Figure 2** (a)Heatmaps (b)Silhouette Plot. Both (a) and (b) are from the same K-means clustering result(k=4)

*Heatmap*

The heatmap is a popular and intuitive visualization technique for encoding quantitative measurements. In the context of gene/microRNA expression data, the color assigned to a cell in the heatmap grid indicates expression value of a particular gene in a given patient sample. The heatmap provides an overview summary of the data. An example is shown in Figure 2(a); the given patient population is stratified into four subgroups. Each subgroup corresponds to one block of the heatmap. The height of each block indicates the number of patients in the corresponding cluster. The main goal of heatmap in our system is to help the user to examine gene expression patterns related to the stratification. In our example, one can observe that the genes show distinct patterns between the clusters while the patterns are consistent within each cluster.

*Silhouette plot*

Clustering results depends heavily on the parameters (e.g., the number of clusters) and distribution of the data points. In order to evaluate the results (and hence the choice of parameters), we incorporated the silhouette plot [24] into our system. The silhouette plot displays the degree of certainty of each sample belonging to its cluster. For our application, this is measured by the difference of a patient's average dissimilarity to other patients of its cluster and the patient's average dissimilarity to all patients to the next closest cluster. The dissimilarity is measured by the squared Euclidean distance between patients in the selected feature space. These dissimilarities are standardized between $-1$ and $1$ and a horizontal bar chart of differences is plotted for each cluster. Thus the silhouette plot allows the user to assess the relative quality of the clusters and provides cues to determine the appropriate number of clusters. Figure 2(b) shows an example of silhouette plot where the number of clusters is 4. The blue and purple clusters are relatively tight and also close to its neighbouring cluster.

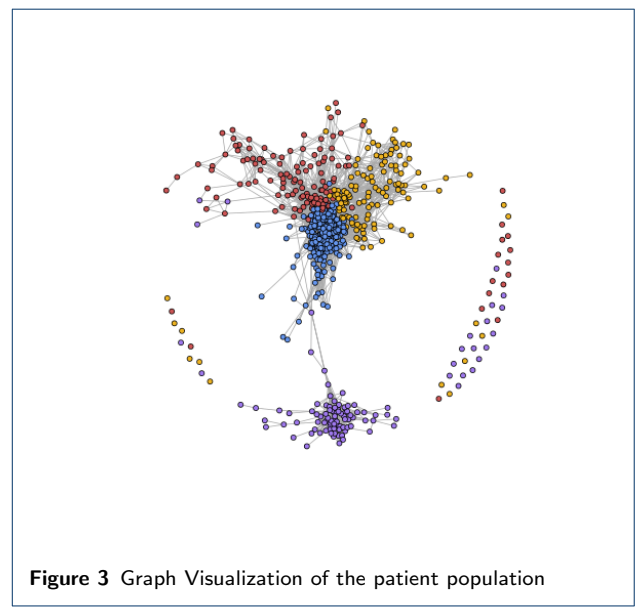

**Figure 3** Graph Visualization of the patient population

*Graph visualization*

Graphs have been widely used to capture relationships among data points. In this work we employ interactive graph visualization of similarity matrices to provide a perspective of the population structure in the feature space and allow users explore patient population. The graph is constructed such that each patient is represented by a vertex, and similarities between patients are represented by edges connecting these vertices. Graph visualizations allows one to browse complex relationships and determine community structures. Generating the graph visualization consists of the following steps:

a) **Similarity Matrix:** The similarity matrix reveals the intrinsic relationships among the entire population. Similarity between two vertices (ie, two patients) can be computed in various ways depending on the property of the measurement data. In our application, the user can choose between Pearson's correlation and the Euclidean distance, two of the most commonly used similarity metrics in bioinformatics.

b) **Graph Sparsification:** The similarity matrix provides a fully connected graph, which needs to be pruned down for effective visualization. The goal of pruning the graph is to extract and summarize the topology of the underlying feature space. We apply a global threshold, keeping only those edges with similarity values exceeding the chosen threshold. With this approach, the resulting graph captures the population structure by connecting only highly similar vertices.

c) **Graph Layout and Vertex Labelling**: Graph layout algorithms project $N$ dimensional data onto two dimensional space for visualization, with the aim of best preserving actual pairwise distances between vertices in the original high dimensional space. We use the force-directed graph layout method of [25] which was shown to be effective in creating uncluttered visualizations.

An example of the graph visualization based on the gene expression features is shown in Figure 3. Since we use a global threshold to sparsify the graph, there are a few isolated nodes. The color labelling of the nodes is consistent with the heatmap and silhouettes plot (Figure 2). It should be observed that several purple and blue nodes are densely connected with nodes of the same color. This confirms our previous observation made with the use of the silhouette plot (see above).

*Parallel Sets*
In order to perform the integrative analysis of different types expression data we employ the parallel sets method. Parallel sets enables users to interactively explore various subgroups obtained as clusters from the clustering step. For each data type, user-selected features (mRNA or microRNAs) are used to separate the patient population into a number of mutually exclusive subgroups by a clustering algorithm. Our application offers $k$-means, spectral clustering and community detection as choices for the clustering algorithm.

Similar to our setting for the heatmap, in parallel sets, patient subgroups are represented as columns. Different types of data are placed independently side-by-side. In each column, subgroups are encoded by boxes whose height is proportional to the number of patients within that subgroup. The color of the boxes indicates the corresponding subgroups, which is consistent with the node color in the graph visualization for the same measurement. The ribbons connecting boxes represent the matching patients in different types of measurements. The width of the ribbons is proportional to the number of patients. The main goal of the ribbons is to offer an intuitive view of the consistency of the clustering result between different measurements. A user can combine multiple subgroups into a larger subgroup by selecting the corresponding regions.

A user can interactively generate a Kaplan-Meier plots [27] by selecting subgroups from the parallel sets visualization. This allows users to evaluate the difference in clinical outcome between selected subgroups. In contrast to the purely data-driven approaches, our application enables the user to treat ribbons as subgroups and can interactively combine subgroups from the clustering algorithm. Thus, a new stratification of the population based on multiple measurements is now obtained. An example is shown in Figure 4.

Implementation
In this section, we describe the workflow design of *iGPSe* and discuss how users interact with the system. Our system workflow has three phases: feature selection, clustering analysis, and integrative patient stratification.

Our system, iGPSe, is developed in Javascript and R using the R/Apache module running on an Apache server. The data processing is implemented with R script which is triggered by Javascript. The interactive visualization part is created using d3.js[28], a visualization JavaScript library.

*Feature selection*
Within the high-dimensional expression data space, each patient is defined by thousands of genes and hundreds of miRNAs. Inevitably, the data sets contain noisy or irrelevant genes/miRNAs which makes subtyping more complicated. Therefore we need to filter out irrelevant genes and only focus on the genes in which the user is interested. iGPSe offers an interactive gene/miRNA list selection panel (Fig. 5). There are three ways to input the feature list: 1)users can select genes/miRNAs of their interest from the table, or 2) type or copy-and-paste the gene names into the input areas or 3) upload a text file (comma delimited format) of the gene list. Once the lists for both genes and miRNAs features are determined, the following patient clustering and stratification will only be based on these selected features. Users can cluster patients in the next step by clicking the 'Next' button.

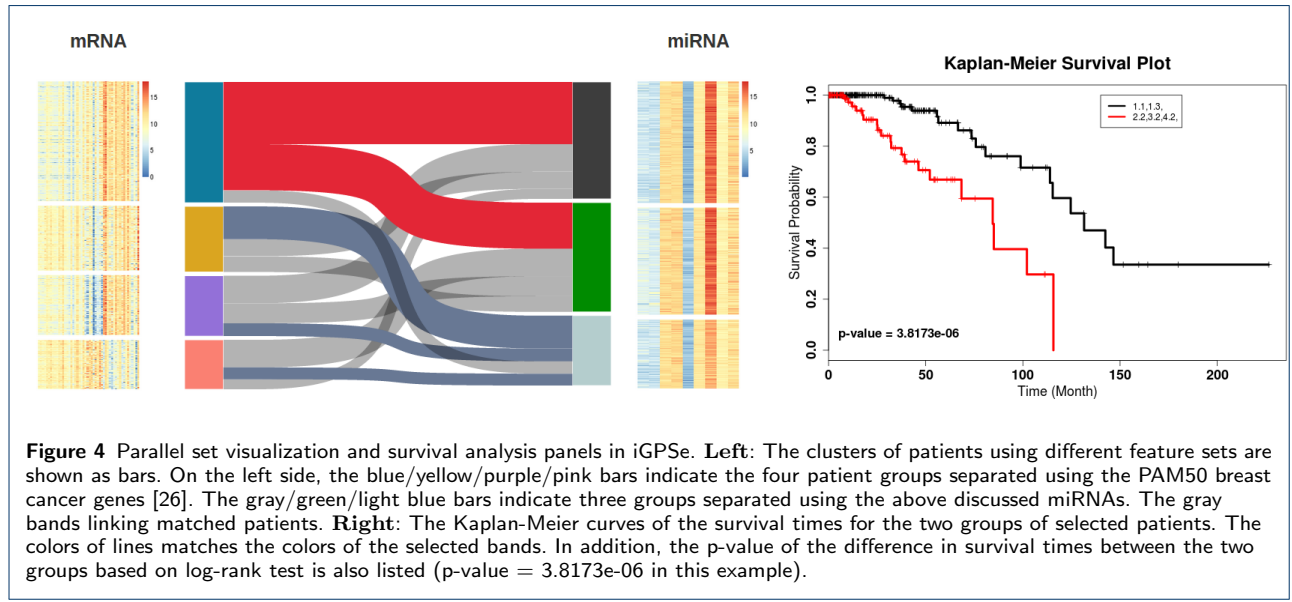

**Figure 4** Parallel set visualization and survival analysis panels in iGPSe. **Left**: The clusters of patients using different feature sets are shown as bars. On the left side, the blue/yellow/purple/pink bars indicate the four patient groups separated using the PAM50 breast cancer genes [26]. The gray/green/light blue bars indicate three groups separated using the above discussed miRNAs. The gray bands linking matched patients. **Right**: The Kaplan-Meier curves of the survival times for the two groups of selected patients. The colors of lines matches the colors of the selected bands. In addition, the p-value of the difference in survival times between the two groups based on log-rank test is also listed (p-value = 3.8173e-06 in this example).

As shown in the Figure 5, the gene list is extracted from a recent *Nature* manuscript describing the TCGA BRCA (breast cancer) project [29]. It has been shown that somatic mutations in these three genes occurred at 10% incidence across all breast cancers, which are the highest among all genes. For the miRNA, we picked hsa-mir-130a, hsa-mir-222, hsa-mir-29a, hsa-mir-23a,hsa-mir-24-1, hsa-mir-24-2, hsa-mir-30a, hsa-mir-27a, hsa-mir-22, and hsa-mir-100 as suggested in [30]. This framework can be easily extended to accommodate all kinds of commonly used high throughput molecular data types and signatures.

*Clustering analysis*
Once the molecular features are selected, iGPSe offers a Clustering section which performs the stratification and help users evaluate and refine the clustering results. The Clustering section provides an interactive interface for tuning parameters of clustering algorithms. We provide users three clustering methods: K-means, spectral clustering, and community detection.

Following the stratification, iGPSe generates the heatmap, silhouette plots and population graph visualization to help user evaluate the validity and quality of the stratification result. iGPSe also provides visualization for an overview of demographic and clinical information such as patients' age distribution and tumorous grades. A Clustering analysis example for the TCGA BRCA dataset using afore-mentioned features is shown in Figure 6.

*Integrative patient stratification*
Integrative patient stratification section allows users to review the clustering results and provides an interactive interface to compare clinical outcomes, such as patient survival and patients ages/tumour grades. Users can select subgroups from the parallel sets to carry out survival analysis including Kaplan-Meier curves and a log-rank test. An example is shown in Figure 1

- **Parallel sets view** shows clustering results of patients using different feature sets which are arranged as columns side-by-side. The columns are split up into disjoint blocks representing clusters. Ribbons connect blocks of two columns, whose width represents how many patients shared between the two clusters.
- **Survival plot view** shows the Kaplan-Meier plots (survival curves) of selected subgroups. If more than one groups is selected, the p-value which tests the null hypothesis that the survival curves are identical in the overall populations using the log-rank test, will be given. Users can select the subgroups by clicking the blocks or ribbons in the parallel sets view.
- **Population graph view** provides an interactive similarity graph visualization. User can drag the graph to adjust the layout, and select a node to acquire correspondence patient's clinical information.

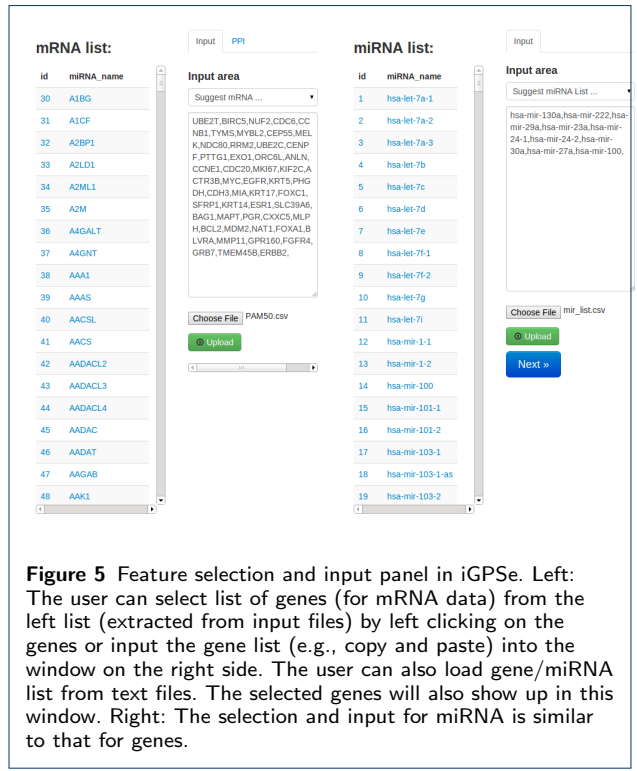

**Figure 5** Feature selection and input panel in iGPSe. Left: The user can select list of genes (for mRNA data) from the left list (extracted from input files) by left clicking on the genes or input the gene list (e.g., copy and paste) into the window on the right side. The user can also load gene/miRNA list from text files. The selected genes will also show up in this window. Right: The selection and input for miRNA is similar to that for genes.

The application also allows users to export the visualization into high quality figures. The users can save a figure to selected format using the 'Save Figures' button. Currently we support PDF, SVG, JPG, and PNG formats.

## Results
Use Case Studies

We demonstrate the use of iGPSe on a patient cohort obtained from the TCGA invasive breast carcinoma

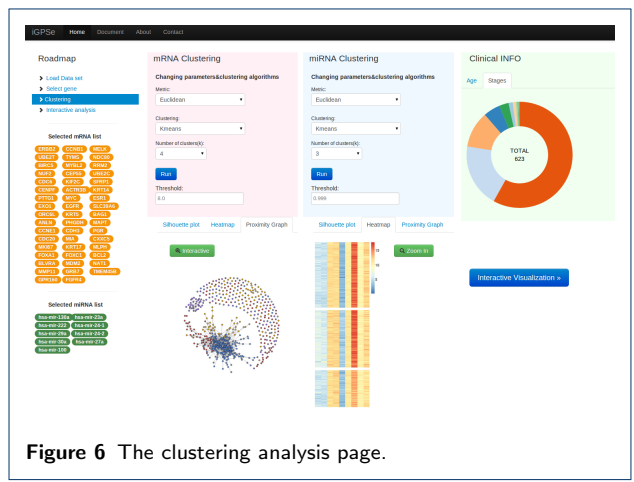

**Figure 6** The clustering analysis page.

(BRCA) project. The TCGA BRCA dataset contains 623 patients' mRNA and miRNA expression profiles as well as clinical outcome information. One of the major advantages of iGPSe is that it enables the user to select genes/miRNAs for use as features to study the patient population.

We are especially interested in the relationship between a group of cell cycle/genome stability genes and the well-known mir-17-92 cluster of miRNAs. In a previous study, we identified a group of more than 400 genes which are frequently co-expressed in multiple types of cancers but not in normal tissues [31]. This gene group is highly enriched with genes involved in cell cycle, mitosis, and genome stability maintenance. It also includes the well-known breast cancer gene panel, PAM 50 [26], which purports to possess prognostic capabilities. The mir-17-92 cluster contains a group of six miRNAs which are proximal on human chromosome 13 and often co-express. They have been shown to be involved in the progression of many types of cancers including lymphoma [32]. Therefore it is of interest to explore if there is any relationship between the frequently co-expressed gene and miRNAs groups in the patient populations and if a combination of them can lead to better prognosis of the breast cancer patients.

Since the data has been log transformed we set Euclidean distance as the similarity function in the preview section. By selecting an appropriate threshold value we get the population graph as in Fig. 1(left graph is constructed from mRNA, right graph is constructed from miRNA). We chose $K$-means as the clustering method to separate patients mRNA and miRNA metric into four and three subgroups, respectively.

Figure 1 shows the parallel view and the network view after the $K$-means clustering of the patients. In the graph visualization, one can note that the population stratifications suggested by the mRNA and miRNA display very different subtyping of patients. In the mRNA graph, patients in the green and brown groups are strongly separated from the orange group, while in the miRNA graph, members in the yellow and grey groups are relatively mixed.

With iGPSe using the parallel sets in the interactive visualization stage, we associate the clustering results with clinical outcomes for the two selected subgroups. We also compare other subgroups from one or more data modalities. Figure 7(a) shows the survival analysis for different choice of subgroups from mRNA only and miRNA only (Figure 7(b)). We can observe that in both mRNA and miRNA exist one subgroup has very different survival times than the others. A similar observation was reported by others [33, 26]. Interestingly, the two patient stratifications are not independent. The blue cluster in mRNA stratification is dominated by samples from grey and green clusters in the miRNA stratification. This observation suggests that combination of mRNA and miRNA features can improve the patient stratification. As shown in Figure 7(c), we can quickly test the choice of other combinations and the result shows a much more significant difference in survival times than just using any single data type to stratify the patient population.

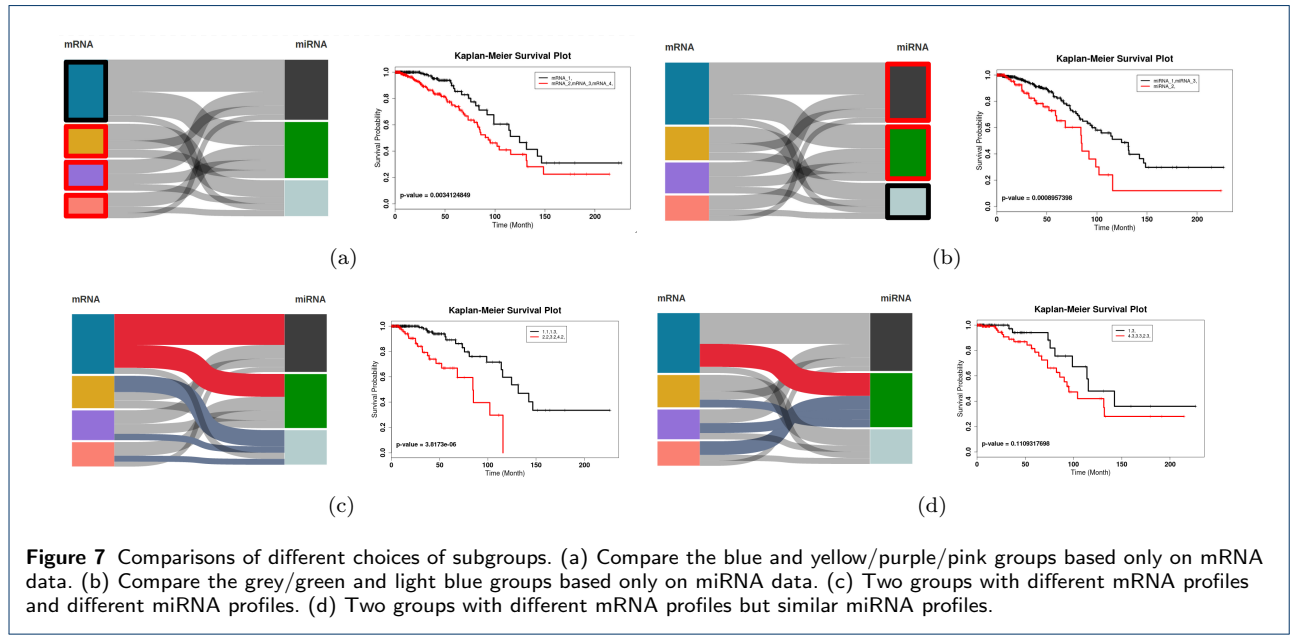

**Figure 7** Comparisons of different choices of subgroups. (a) Compare the blue and yellow/purple/pink groups based only on mRNA data. (b) Compare the grey/green and light blue groups based only on miRNA data. (c) Two groups with different mRNA profiles and different miRNA profiles. (d) Two groups with different mRNA profiles but similar miRNA profiles.

Users' Feedback

We collected comments from three domain experts (Biostatistician, clinician and basic scientist) on the use of iGPSe, providing some evidence that our visualization and interaction design, grounded in a characterization of the domain requirements, supported efficient explorations of subtypes. We prepared datasets for the TCGA breast invasive carcinoma (BRCA) cohorts and invited our domain experts to perform population stratification using iGPSe. The initial user feedbacks on the utility of iGPSe are positive.

Before we present the feedbacks and evaluation details, we outline typical analysis workflow in genomics based cancer patient stratification. As we stated earlier, a standard genomic based cancer patient stratification workflow includes steps of data curation and pre-processing, clustering, subtype characterization and validation. Traditionally, these steps involve scripts written in different platforms which requires researchers possess at least a basic level of programming skills. Using separate static scripts makes the analysis of large populations with multiple modalities laborious, in particular when studying the interactions between different data types. The design of iGPSe was directly motivated by these shortcomings and the following observations reflect how domain scientists used iGPSe to analyze the model described in previous section.

- **Feature Selection** – The identification of discriminant genes is of fundamental and practical interest. A small subset of highly discriminant genes can be extracted to build more reliable cancer classifiers. For users who do not have experience in writing scripts to process the input data, iGPSe interactive GUI offers a more convenient way than writing scripts to select the target genes list. A collaborator mentioned that scroll down table of genes and input genes can help users adding and dropping genes from the selected list effectively, especially for people who do not have a programming background.
- **Clustering section** – The general aim of the clustering section is clustering patient population into subgroups and examine the results' quality for each data type. Our collaborators agreed that the interactive clustering parameters control panel allows users who have no prior knowledge of machine learning to guide the stratification process. In addition, the graph visualization of the patient population gives an overview of the population density in the selected feature space which helps users to verify the clustering results are consistent with the proximity graph. The silhouette plot of the stratification gives a rapid overview of the quality of the clustering result. This immediate visual access makes the judgment of simulation output faster and more intuitive. Moreover, all views can be compared across runs simply by launching the visualization multiple times. The traditional workflow would have forced scientists to manually run scripts.

- **Integrative analysis** – our collaborators noted that parallel set visualization helps them to explore the relationship between stratification results from two different types of data. They also told us that the interactive combination of two types of data stratification results with the parallel sets is very useful, especially interactively generating the survival plot. Eventually experts are able to answer more challenging questions such as whether patient groups which share similar gene expression have very different clinical outcomes when they are expressed differently with the miRNA.

In general, very positive outcome of the evaluation sessions with our collaborators was that in all cases they asked us to load further data to explore with iGPSe. This is convenient especially for testing or replicating reported gene lists.

## Conclusion

Visual analytics is an emerging discipline that combines visualization methods with data analysis and human-computer interaction. As shown in this paper, application of visual analytics methods in integrative genomics can enable quick integration of different types of data and significantly facilitate the discovery of integrated molecular markers for cancer subtyping and outcome prediction.

In our case study, it is observed that the two groups of patients with different gene (mRNA) expression profiles for the previously identified genome stability genes show differences in survival times only when they have similar specific expression profiles for the miRNAs in the mir-17-92 cluster. This observation suggests that the genome stability genes and mir-17-92 cluster may influence breast cancer development and progression in different pathways even though the genes such as PAM50 and mir-17-92 cluster interact with each other. Thus it is of great interest to study the genes and pathways targeted by the mir-17-92 clusters in order to elucidate the different mechanisms.

While we have demonstrated the functionalities of iGPSe using only mRNA and miRNA signatures, it nevertheless can accommodate other types of patient information such as genome-specific information, DNA methylation, and even morphological features extracted from pathological images. In addition, the system can accommodate comparison among more than two subgroups as well as more than two types of data, which makes iGPSe highly versatile for biomedical researchers to use and generate highly interpretable results much more promptly than the cumbersome script-based approach.

Our evaluation with domain experts shows that major strengths of iGPSe is that it eliminated the needs of programming and scripting from users while still grant users sufficient control during the steps including feature selection, clustering, subgroup selection and comparison. The automatic comparison of clinical outcome (i.e., survival) is of particular interests to the users.

In the future, we plan to make this a publicly accessible web tool as it is currently implemented using an Apache server. Users will be able to upload and analyze their own data with iGPSe. In addition, more choices on the clustering algorithms will be implemented. From the machine learning point of view, our approach provides an alternative way of carrying out consensus clustering in order to reconcile different clustering results from different features. This system can be combined with existing consensus clustering approaches to further streamline the subgroup selection process. Since graph visualization allows interactive manipulation of the data points, a feedback mechanism for interactively assigning clustering membership based on visualization can be deployed to enable iterative feature selection.

Overall, we have demonstrated that by combining graph visualization with parallel views and bioinformatics analysis, we can significantly reduce the computing burden for biomedical researchers in order to explore the complicated integrated genomics data. This approach is generalizable to enable more sophisticated analysis for cancer biomarker discovery and subtyping in order to achieve precision medicine.

**Competing interests**
The authors declare that they have no competing interests.

**Author's contributions**
HD carried out the system design and implementation, and drafted the manuscript. CW participated in the system design and coordination of case studies. KH and RM conceived of the study, designed the iGPSe, and contributed to discussions and suggestions to complete the manuscript. RM supervised the project. All authors read and approved the final manuscript.

**Acknowledgements**
This work is partially supported by National Institutes of Health under Grant R01CA141090.

**Author details**
[1]Computer Science& Engineering Department, The Ohio State University, 43210 Columbus, OH, US. [2]Biomedical Informatics Department, The Ohio State University, 43210 Columbus, OH, US. [3]Electrical & Computer Engineering, 43210 Columbus, OH, US.